\documentclass{PoS}

\def\simgt{\rlap{\lower 3.5 pt\hbox{$\mathchar \sim$}}\raise 1.0pt \hbox {$>$}}

\title{Systematic study of operator dependence in nucleus calculation at large quark mass}

\ShortTitle{Systematic study of operator dependence in nucleus calculation at large quark mass}

\author{
   \speaker{Takeshi Yamazaki}$^{a,b,c}$,  
   Ken-Ichi Ishikawa$^{d,c}$,
   Yoshinobu Kuramashi$^{b,a,c}$,
   Akira~Ukawa$^{c}$
   \ for PACS Collaboration\\
   \\
   \\
   \llap{$^a$}
Faculty of Pure and Applied Sciences,
University of Tsukuba, Tsukuba, Ibaraki 305-8571, Japan
   \\
   \llap{$^b$}
Center for Computational Sciences, University of Tsukuba,
Tsukuba, Ibaraki 305-8577, Japan
   \\
   \llap{$^c$}
RIKEN Advanced Institute for Computational Science,
Kobe, Hyogo 650-0047, Japan
   \\
   \llap{$^d$}
Department of Physics, Hiroshima University, Higashi-Hiroshima, Hiroshima 739-8526, Japan.\\
Email: \email{yamazaki@het.ph.tsukuba.ac.jp}
}

\abstract{Recently it is claimed that there is a significant systematic 
error from 
excited state contributions in the nucleus correlation functions by
comparing with calculations using
the exponential and wall source operators. However, the wall source result 
is obtained in much earlier time than the plateau region.
In order to investigate the systematic error in the plateau region, 
we calculate the correlation functions with both the operators
in quenched QCD at 0.8 GeV pion mass and in $N_f=2+1$ QCD at 0.7 GeV pion mass
in high accuracy. 
In this report we present preliminary results of those calculations, 
and show that the energy shift obtained from the two sources
agree with each other, if those are determined from
a region, where both the nucleon and two-nucleon correlation
functions have plateaus.
}

\FullConference{34th annual International Symposium on Lattice Field Theory\\
		24-30 July 2016\\
		University of Southampton, UK}

\begin{document}

\section{Introduction}

The binding energies of light nuclei with the mass number less than or equal to
four were calculated in lattice QCD
by three groups~\cite{Yamazaki:2009ua,Yamazaki:2011nd,Beane:2011iw,Beane:2012vq,Yamazaki:2012hi,Berkowitz:2015eaa,Yamazaki:2015asa,Orginos:2015aya}.
Those results were obtained from the exponential or gaussian sources.
Recently, HALQCD~\cite{Iritani:2016jie} claimed a possibility of large 
systematic error coming from excited state contaminations, in
other words source operator dependence, in the binding energy calculation.
Therefore, in this study, we investigate source dependence of 
the energy shift in the spin-triplet two-nucleon channel 
using the exponential and wall sources.

In the binding energy calculation, the correlation functions for
the nucleon $C_N(t)$ and two-nucleon $C_{NN}(t)$ channels are calculated,
and the energy shift between the twice of the nucleon mass and
the ground state two-nucleon energy, $2m_N - E_{NN}$, is evaluated from
the ratio of the correlation functions, $C_{NN}(t)/(C_N(t))^2$.
An important condition of this analysis is that the energy shift
must be determined from plateau regions of $C_N(t)$ and $C_{NN}(t)$.
It is known that the wall source needs much larger $t$ 
than other smearing sources to obtain
the ground state energy for the nucleon and multi-nucleon systems.
Thus, we employ large quark masses, corresponding to $m_\pi = 0.8$ GeV 
in $N_f = 0$ and $0.7$ GeV in $N_f = 2 + 1$, 
and also carry out huge number of measurement,
more than $10^5$, of the correlation functions
to obtain a signal in large $t$ region in the wall source calculation.
Using the data in high accuracy for the exponential and wall sources,
we study $t$ dependence of the effective energy shift in each source,
and compare results for the two sources in each plateau region
to examine whether these sources give different results or not.
We also investigate volume dependence of the energy shift obtained
from the wall source.
All the results in this report are preliminary.

\section{Simulation parameters}

For the quenched QCD configuration, we employ Iwasaki gauge action
at $\beta = 2.416$, corresponding to $a = 0.128$ fm~\cite{AliKhan:2001tx}.
The quark propagators are calculated with a tad-pole improved Wilson
action with $c_{\rm SW} = 1.378$ at $\kappa_{ud} = 0.13482$ corresponding
to $m_\pi = 0.8$ GeV.
The actions and parameters are the same as in 
our previous works~\cite{Yamazaki:2009ua,Yamazaki:2011nd}.
The temporal lattice size is fixed to 64, while the spatial size is
chosen to be 16, 20, and 32.

For the $N_f = 2+1$ QCD configuration, we use 
Iwasaki gauge and a nonperturbative $O(a)$ improved Wilson actions
with the same parameters of $\beta$ and $c_{\rm SW}$ 
in Ref.~\cite{Aoki:2009ix}, where $a = 0.09$ fm.
We employ $\kappa_{ud} = 0.1369425$ and $\kappa_s = 0.1368530$
corresponding to $m_\pi = 0.7$ GeV.
The lattice size is $L^3 \times T = 32^3 \times 48$.

We calculate the correlation functions for the nucleon and two-nucleon
spin-triplet channel using the exponential and wall sources with the point sink.
In order to increase statistics, 
we carry out various measurements of the correlation functions
in each configuration by
changing the source spacetime position 
and source time slice for the exponential and 
wall sources, respectively.

\section{Result}

\subsection{$N_f = 0$ QCD at $m_\pi = 0.8$ GeV}

\begin{figure}[!t]
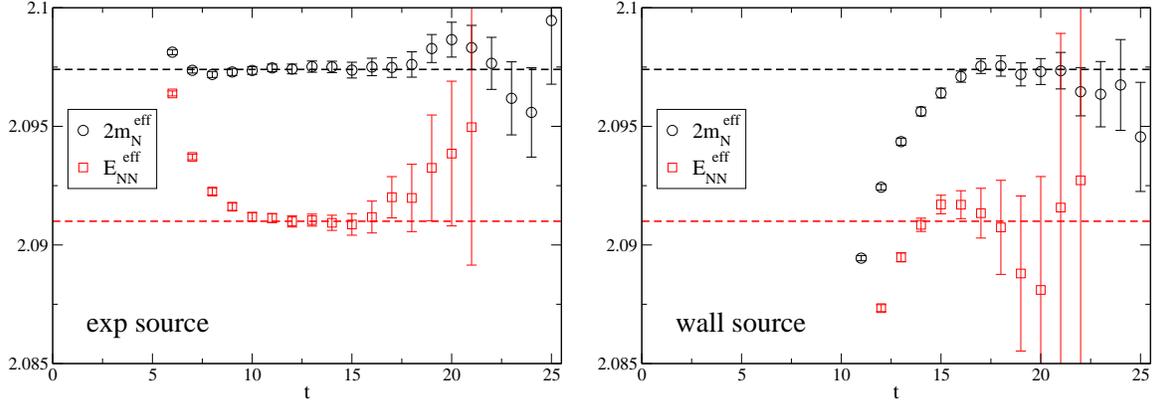

\begin{tabular}{cc}
\includegraphics*[angle=0,width=0.49\textwidth]{Figs/Nf0/eff_NN31whl.eps}
&
\includegraphics*[angle=0,width=0.49\textwidth]{Figs/Nf0/w_eff_NN31whl.eps}
\end{tabular}
\caption{
Effective nucleon mass $m_N^{\rm eff}$ (circle) and 
two-nucleon energy $E_{NN}^{\rm eff}$ (diamond) in the spin-triplet channel
using the exponential (left panel) and wall (right panel) sources
in the quenched case at $m_\pi = 0.8$ GeV.
To compare $m_N^{\rm eff}$ with $E_{NN}^{\rm eff}$,
$m_N^{\rm eff}$ is multiplied by 2 in the figures.
The dashed lines express the values of each plateau in the 
exponential source.
\label{fig:nf0:comp}
}
\end{figure}

\begin{figure}[!t]
\hfil
\includegraphics*[angle=0,width=0.49\textwidth]{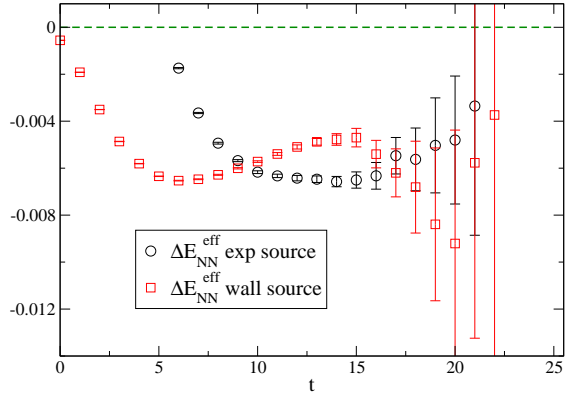}
\caption{
Effective energy shift 
$\Delta E_{NN}^{\rm eff} = 2m_N^{\rm eff}-E_{NN}^{\rm eff}$
using the exponential (circle) and wall (diamond) sources
in the quenched case at $m_\pi = 0.8$ GeV.
\label{fig:nf0:comp:dE}
}
\end{figure}

Figure~\ref{fig:nf0:comp} presents twice of the effective nucleon mass and
the effective two-nucleon energy with the exponential (left panel)
and wall (right panel) sources on $L=20$.
The number of the measurement is about 5$\times 10^6$ in both the sources.
The results of the exponential source have reasonable plateaus, which start
from $t \sim 11$.
The dashed line expresses each value in the plateau region of the exponential
source.
The results of the wall source have larger errors in the plateau region
than the exponential one due to the late plateau, which start from $t\sim 16$.
The effective mass and energy of the wall source in the plateau region 
are reasonably 
consistent with the ones obtained from the exponential source.

The effective energy shift, which is evaluated from the ratio of
the correlation functions $C_{NN}(t)/(C_N(t))^2$,
is plotted in Fig.~\ref{fig:nf0:comp:dE}.
A reasonable plateau is seen in the result of the exponential source,
whose region is consistent with the plateau region of each correlation
function shown in the left panel of Fig.~\ref{fig:nf0:comp}.
On the other hand, the result of the wall source does not have
a plateau in $t < 15$.
The reason is that both the correlation functions in the ratio
do not reach plateaus in this region as shown
in the right panel of Fig.~\ref{fig:nf0:comp}.
In the plateau region of each correlation function, $t \simgt 16$,
the result of the wall source agrees with the plateau value of the
exponential source within the large error.

From the comparison, we conclude that a consistent energy
shift is obtained from the exponential and wall sources, when
those values are determined from each plateau region.
Furthermore, the wall source needs much larger statistics than 
the exponential source to obtain a clear signal of the energy shift.

\begin{figure}[!t]
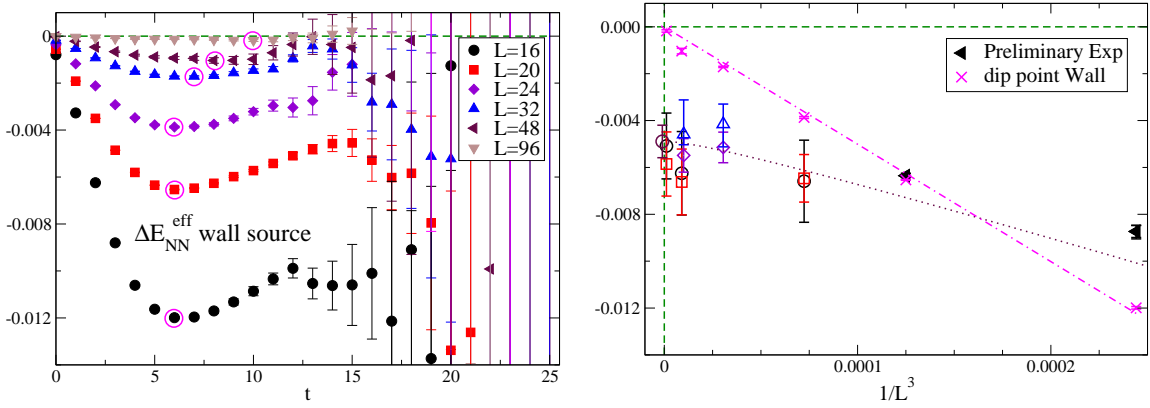

\begin{tabular}{cc}
\includegraphics*[angle=0,width=0.49\textwidth]{Figs/Nf0_W_Vdep/dE_NN31wcl.eps}
&
\includegraphics*[angle=0,width=0.49\textwidth]{Figs/Nf0_W_Vdep/DdE_31wW.eps}
\end{tabular}
\caption{
The effective energy shift 
$\Delta E_{NN}^{\rm eff}$ using the wall source 
on several volumes (left panel) and 
volume dependence of the energy shift (right panel).
In the left panel, the data on $L=16, 20, 32$ are preliminary results
in the current calculation, while those on $L=24, 48, 96$ are the ones
in pilot studies for the previous work~\cite{Yamazaki:2011nd}.
In the right panel, the cross symbols are the data of the dip point
expressed by the open circle in the left panel, and the dash-dot line
is for the guide of eye.
The open symbols and dashed line are the results in 
the previous works~\cite{Yamazaki:2011nd}.
The closed triangle represents a preliminary result obtained from
the exponential source in the current calculation.
\label{fig:nf0:vdep}
}
\end{figure}

The volume dependence of the effective energy shift using the wall
source is plotted in the left panel of Fig.~\ref{fig:nf0:vdep},
which includes the data of this work ($L=16, 20, 32$) and also
the ones of a pilot study of
the previous work~\cite{Yamazaki:2011nd} ($L=24, 48, 96$).
In small $t$ region the effective energy shift largely depends on the volume, 
and increases with the volume.
The results in all the volumes have a dip in the region.
If statistics is not enough, one might regard the dip as a plateau,
especially in larger volume, because the $t$ dependence is milder.
When we pick up the dip point in each volume, which is the minimum value in the
small $t$ region as expressed by open circles in the figure, 
they roughly behave as $1/L^3$ as presented
in the right panel of Fig.~\ref{fig:nf0:vdep}.
This behavior is consistent with the one observed by HALQCD
using the wall source in Ref.~\cite{Iritani:2016jie}.
This, however, is not the volume dependence of the ground state energy shift,
because it must be determined from the plateau region,
which is roughly more than 15 in our wall source data
as shown in Fig.~\ref{fig:nf0:comp}.
In our data, the effective energy shift has large error in the region,
and it is hard to estimate a precise energy shift from the wall source data.
On the other hand, from the exponential source 
the ground state energy shifts can be determined precisely
as shown in the right panel of 
Fig.~\ref{fig:nf0:vdep}, whose volume dependence is reasonably consistent with 
the one of the previous work~\cite{Yamazaki:2011nd}.

\subsection{$N_f = 2+1$ QCD at $m_\pi = 0.7$ GeV}

\begin{figure}[!t]
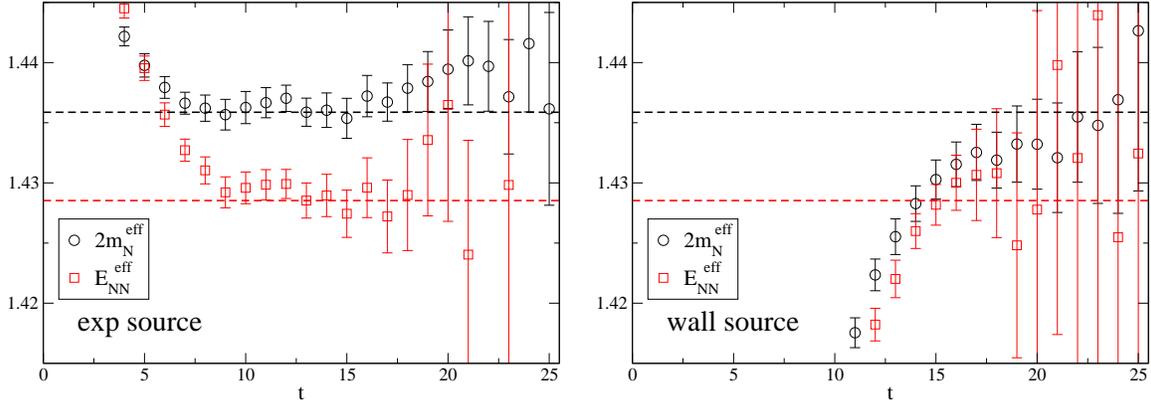

\begin{tabular}{cc}
\includegraphics*[angle=0,width=0.49\textwidth]{Figs/Nf2+1/eff_NN31whl.eps}
&
\includegraphics*[angle=0,width=0.49\textwidth]{Figs/Nf2+1/w_eff_NN31whl.eps}
\end{tabular}
\caption{
The same figure as Fig.~\protect\ref{fig:nf0:comp}, but for
$N_f = 2 + 1$ calculation at $m_\pi = 0.7$ GeV.
\label{fig:nf2+1:comp}
}
\end{figure}

\begin{figure}[!t]
\hfil
\includegraphics*[angle=0,width=0.49\textwidth]{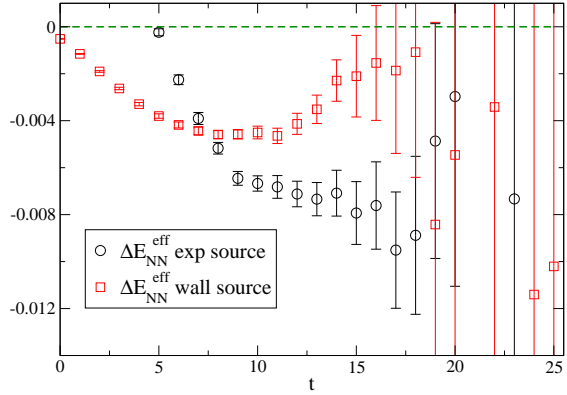}
\caption{
The same figure as Fig.~\protect\ref{fig:nf0:comp:dE}, but for
$N_f = 2 + 1$ calculation at $m_\pi = 0.7$ GeV.
\label{fig:nf2+1:comp:dE}
}
\end{figure}

Figure~\ref{fig:nf2+1:comp} is the same figure as Fig.~\ref{fig:nf0:comp},
but for the $N_f = 2 + 1$ calculation.
The statistics is smaller than the ones in the quenched case.
The numbers of the measurement for the exponential and wall sources
are $2 \times 10^5$ and $1 \times 10^5$, respectively.
As in the quenched case, the data of the exponential source have
plateau, which start from $t \sim 9$. 
For the wall source, it is unclear where plateaus start due to
larger statistical error.
The effective nucleon mass seems to have a plateau in the region $t \simgt 16$,
while the value of the effective mass is smaller than the one of
the exponential source.
When the statistics is increased, the plateau of the wall source will
appear in a larger $t$ region, $t \simgt 20$, 
and its value will agree with the result of the exponential source.

The effective energy shifts from the two sources are plotted in
Fig.~\ref{fig:nf2+1:comp:dE}.
The result of the exponential source has a reasonable plateau in the
region, where the effective mass and energy have their plateau as 
shown in the left panel of Fig.~\ref{fig:nf2+1:comp}.
In the wall source a $t$ independent behavior is seen 
in $t = 8$--11,
although the effective mass and energy do not 
reach a plateau in this region.
This is a similar behavior to the dip structure as seen in the quenched data.
Since the lattice spacing in $N_f = 2+1$ is finer than
the one in the quenched case, the dip structure becomes wider in the
$N_f = 2 + 1$ case. 
The $t$ independent behavior would be regarded as a plateau, if
the statistics is not enough. 
In such a case, one might conclude that different results are obtained
from the two sources.
However, the energy shift of the ground state must be determined from
the plateau region of each correlator in 
the right panel of Fig.~\ref{fig:nf2+1:comp}.
Unfortunately, our data do not have enough statistics to obtain
the effective energy shift clearly in this region,
while the value in $t \simgt 20$ is consistent with the one obtained 
from the exponential source within the large error of the wall source.
In order to check the consistency further,
it is important to increase statistics by generating
configurations.

\section{Summary}

We have investigated source operator dependence of
the energy shift in the spin-triplet two-nucleon channel
in the $N_f = 0$ and $N_f = 2 + 1$ cases.
From the comparisons with the results with the exponential
and wall sources, it is concluded that the two sources 
give consistent results in each plateau region, 
when the statistics is large enough.
Using the exponential source, the energy shift can be determined from
the plateau regions for the nucleon and two-nucleon correlation functions,
while from the wall source it is hard to obtain the energy shift 
in a similar quality to the exponential source due to the late plateaus
for the correlation functions.

It is also concluded that the effective energy shift in the wall
source always has the dip structure in the small $t$ region, where
the nucleon and two-nucleon correlation functions do not reach 
a plateau.
If the statistics is not enough, the dip structure might look like
a plateau, especially in a fine lattice spacing on a large volume.
In such a case, the energy shift, which is determined from the dip point,
roughly behaves as $1/L^3$.
This is a consistent behavior with the wall source result obtained by
HALQCD Collaboration in Ref.~\cite{Iritani:2016jie}.

We consider that the properties observed in the current study 
at the large $m_\pi$
will be seen in smaller $m_\pi$ calculation, 
because in our previous 
works~\cite{Yamazaki:2009ua,Yamazaki:2011nd,Yamazaki:2012hi,Yamazaki:2015asa}
qualitative differences, such as for the existence of the light nuclei,
were not observed in $m_\pi = 0.3$-0.8 GeV.
Although the consistency between the exponential and wall sources
is investigated in this calculation, more reliable results can
be obtained from the method using the generalized eigenvalue problem
~\cite{Luscher:1990c}.
Therefore, it is an important future work to calculate the two-nucleon
energy using the more reliable method, and compare the energy of the 
ground state using the method with those obtained from plateau regions of 
the exponential and wall sources.

\section*{Acknowledgements}
Numerical calculations for the present work have been carried out on 
the FX10 supercomputer system at Information Technology Center of 
the University of Tokyo, on the COMA cluster system under 
the ``Interdisciplinary Computational Science Program'' of 
Center for Computational Science at University of Tsukuba, 
on the Oakforest-PACS system of Joint Center for 
Advanced High Performance Computing,
on the computer facilities of the Research Institute for 
Information Technology of Kyushu University, and on
the FX100 and CX400 supercomputer systems at the Information Technology Center 
of Nagoya University.
This research used computational resources of the HPCI system provided by 
Information Technology Center of the University of Tokyo through 
the HPCI System Research Project (Project ID: hp160125). 
We thank the colleagues in the PACS Collaboration for providing us 
the code used in this work. This work is supported in part by Grants-in-Aid 
for Scientific Research from the Ministry of Education, Culture, Sports, 
Science and Technology (Nos. 25800138, 16H06002).

\bibliography{lat16}

\providecommand{\href}[2]{#2}\begingroup\raggedright\begin{thebibliography}{10}

\bibitem{Yamazaki:2009ua}
{\bf PACS-CS Collaboration}, T.~Yamazaki, Y.~Kuramashi, and A.~Ukawa, {\em
  Phys.Rev.} {\bf D81} (2010) 111504.

\bibitem{Yamazaki:2011nd}
{\bf PACS-CS Collaboration}, T.~Yamazaki, Y.~Kuramashi, and A.~Ukawa, {\em
  Phys. Rev.} {\bf D84} (2011) 054506.

\bibitem{Beane:2011iw}
{\bf NPLQCD Collaboration}, S.~Beane et~al., {\em Phys. Rev.} {\bf D85} (2012)
  054511.

\bibitem{Beane:2012vq}
{\bf NPLQCD Collaboration}, S.~Beane, E.~Chang, S.~Cohen, W.~Detmold, H.~Lin,
  et~al., {\em Phys.Rev.} {\bf D87} (2013), no.~3 034506.

\bibitem{Yamazaki:2012hi}
T.~Yamazaki, K.-i. Ishikawa, Y.~Kuramashi, and A.~Ukawa, {\em Phys.Rev.} {\bf
  D86} (2012) 074514.

\bibitem{Berkowitz:2015eaa}
{\bf CalLat Collaboration}, E.~Berkowitz, T.~Kurth, A.~Nicholson, B.~Joo,
  E.~Rinaldi, M.~Strother, P.~M. Vranas, and A.~Walker-Loud, {\em Phys. Lett.}
  {\bf B765} (2017) 285--292.

\bibitem{Yamazaki:2015asa}
T.~Yamazaki, K.-i. Ishikawa, Y.~Kuramashi, and A.~Ukawa, {\em Phys. Rev.} {\bf
  D92} (2015), no.~1 014501.

\bibitem{Orginos:2015aya}
{\bf NPLQCD Collaboration}, K.~Orginos, A.~Parreno, M.~J. Savage, S.~R. Beane,
  E.~Chang, and W.~Detmold, {\em Phys. Rev.} {\bf D92} (2015), no.~11 114512.

\bibitem{Iritani:2016jie}
{\bf HALQCD Collaboration}, T.~Iritani et~al., {\em JHEP} {\bf 10} (2016) 101.

\bibitem{AliKhan:2001tx}
{\bf CP-PACS Collaboration}, A.~Ali~Khan et~al., {\em Phys. Rev.} {\bf D65}
  (2002) 054505.

\bibitem{Aoki:2009ix}
{\bf PACS-CS Collaboration}, S.~Aoki et~al., {\em Phys. Rev.} {\bf D81} (2010)
  074503.

\bibitem{Luscher:1990c}
M.~L{\"u}scher and U.~Wolff, {\em Nucl. Phys.} {\bf B339} (1990) 222--252.

\end{thebibliography}\endgroup

\end{document}